\documentclass[sigconf,screen]{acmart}

\AtBeginDocument{%
  \providecommand\BibTeX{{%
    \normalfont B\kern-0.5em{\scshape i\kern-0.25em b}\kern-0.8em\TeX}}}

\setcopyright{acmlicensed}
\acmDOI{10.1145/3663529.3663796}
\acmYear{2024}
\copyrightyear{2024}
\acmSubmissionID{fsecomp24demo-p3-p}
\acmISBN{979-8-4007-0658-5/24/07}
\acmConference[FSE Companion '24]{Companion Proceedings of the 32nd ACM International Conference on the Foundations of Software Engineering}{July 15--19, 2024}{Porto de Galinhas, Brazil}
\acmBooktitle{Companion Proceedings of the 32nd ACM International Conference on the Foundations of Software Engineering (FSE Companion '24), July 15--19, 2024, Porto de Galinhas, Brazil}
\received{2024-01-29}
\received[accepted]{2024-04-15}

\usepackage{xspace}
\usepackage{comment}
\usepackage{xcolor, soul}
\usepackage{xcolor,colortbl}
\usepackage{enumitem}
\usepackage{graphicx}
\usepackage{hyperref}
\usepackage{amsmath}
\usepackage{verbatim}
\usepackage{multirow}

\usepackage{xcolor}
\usepackage{textcase}
\usepackage{caption}
\usepackage{tabularx}
\usepackage{tabularray}
\UseTblrLibrary{booktabs}
\usepackage{makecell}
\usepackage{booktabs}
\usepackage{siunitx}
\usepackage[utf8]{inputenc}
\usepackage[T1]{fontenc}
\usepackage{microtype}
\usepackage{flushend}

\definecolor{dr}{HTML}{8b0000}

\usepackage{bbm}

\usepackage{svg}
\usepackage{booktabs}
\usepackage{array}
\usepackage{diagbox}
\newcommand{\PreserveBackslash}[1]{\let\temp=\\#1\let\\=\temp}
\newcolumntype{C}[1]{>{\PreserveBackslash\centering}p{#1}}
\newcolumntype{R}[1]{>{\PreserveBackslash\raggedleft}p{#1}}
\newcolumntype{L}[1]{>{\PreserveBackslash\raggedright}p{#1}}

\usepackage{MnSymbol}
\definecolor{Gray}{gray}{0.9}
\definecolor{light_green}{HTML}{c4edc9}
\definecolor{light_red}{HTML}{fddad9}
\definecolor{light_blue}{HTML}{cad7fb}
\definecolor{light_yellow}{HTML}{FFE599}
\definecolor{lightyellow}{HTML}{ffffed}
\definecolor{lightpurple}{HTML}{E6E6FA}

\DeclareRobustCommand{\hlgreen}[1]{{\sethlcolor{light_green}\hl{#1}}}

\definecolor{tab_highlight}{rgb}{0.99,0.80,0.49}

\newcolumntype{N}[1]{>{\centering\arraybackslash}m{#1cm}}

\usepackage{color, colortbl}

\newcounter{finding}

\usepackage{tcolorbox}

\usepackage{subcaption}

\usepackage{wrapfig}

\usepackage{listings}
\usepackage{xcolor}

\definecolor{codegreen}{rgb}{0,0.6,0}
\definecolor{codegray}{rgb}{0.5,0.5,0.5}
\definecolor{codepurple}{rgb}{0.58,0,0.82}
\definecolor{backcolour}{rgb}{0.95,0.95,0.92}

\definecolor{darkBlue}{rgb}{0.000000,0.000000,0.545098}
\definecolor{darkGreen}{rgb}{0.000000,0.392157,0.000000}
\definecolor{DarkGray}{gray}{0.4}
\definecolor{javared}{rgb}{0.6,0,0} 
\definecolor{javagreen}{rgb}{0.25,0.5,0.35} 
\definecolor{javapurple}{rgb}{0.5,0,0.35} 
\definecolor{javadocblue}{rgb}{0.25,0.35,0.75} 
\definecolor{lightgray}{gray}{0.8}
\definecolor{lightblue}{rgb}{0.63, 0.79, 0.95}

\definecolor{shadecolor}{RGB}{150,150,150}
\definecolor{blueA}{RGB}{204,229,255}
\definecolor{redA}{RGB}{112,0, 0}
\definecolor{RED}{RGB}{255,0, 0}

\lstdefinestyle{mystyle}{
    frame=none,
  xleftmargin=15pt,
  stepnumber=1,
  numbers=left,
  numbersep=5pt,
  stepnumber=1,
  numberstyle=\tiny\bf,
  belowcaptionskip=\bigskipamount,
  captionpos=b,
  escapeinside={*‘}{’*},
  tabsize=5,
  emphstyle={\bf},
  basicstyle=\scriptsize\ttfamily,
  keywordstyle=\color{javapurple}\bfseries,
  stringstyle=\color{javared},
  commentstyle=\color{javagreen},
  morecomment=[s][\color{javadocblue}]{/**}{*/},
  showspaces=false,
  columns=flexible,
  showstringspaces=false,
  morecomment=[l]{//},
  tabsize=2,
  breaklines=true
}

\definecolor{light_gray}{HTML}{d5d5d5}

\DeclareRobustCommand{\hlgray}[1]{{\sethlcolor{light_gray}\hl{#1}}}

\newcommand{\tool}{\textsc{Decide}}
\newcommand{\webtool}{\textsc{Decide}}

\begin{document}

\title{Decide: Knowledge-Based Version Incompatibility Detection in Deep Learning Stacks}

\author{Zihan Zhou}
\orcid{0009-0007-2817-3113}
\affiliation{%
  \institution{The University of Hong Kong}
  \city{Hong Kong}
  \country{Hong Kong}
}
\email{zihan2@connect.hku.hk}

\author{Zhongkai Zhao}
\orcid{0000-0003-2365-9898}
\affiliation{%
  \institution{National University of Singapore}
  \city{Singapore}
  \country{Singapore}
}
\email{zhongkai.zhao@u.nus.edu}

\author{Bonan Kou}
\orcid{0000-0003-1407-8522}
\affiliation{%
  \institution{Purdue University}
  \city{West Lafayette}
  \country{USA}
}
\email{koub@purdue.edu}

\author{Tianyi Zhang}
\orcid{0000-0002-5468-9347}
\affiliation{%
  \institution{Purdue University}
  \city{West Lafayette}
  \country{USA}
}
\email{tianyi@purdue.edu}
\renewcommand{\shortauthors}{Zihan Zhou, Zhongkai Zhao, Bonan Kou and Tianyi Zhang}

\begin{abstract}
Version incompatibility issues are prevalent when reusing or reproducing deep learning (DL) models and applications. 
Compared with official API documentation, which is often incomplete or out-of-date, Stack Overflow (SO) discussions possess a wealth of version knowledge that has not been explored by previous approaches.
To bridge this gap, we present \webtool{}, a web-based visualization of a knowledge graph that contains 2,376 version knowledge extracted from SO discussions.
As an interactive tool, \webtool{} allows users to easily check whether two libraries are compatible and explore compatibility knowledge of certain DL stack components with or without the version specified.
A video demonstrating the usage of \webtool{} is available at \url{https://youtu.be/wqPxF2ZaZo0}.
\end{abstract}



\maketitle

\section{Introduction}

Deep learning (DL) has been widely applied in diverse domains, such as computer vision~\cite{wu2017application}, natural language processing~\cite{young2018recent}, and autonomous driving~\cite{tian2018deeptest}.
However, most DL applications are built on a complex and heterogeneous DL stack~\cite{huang2022demystifying}, including libraries, runtime systems, drivers, operating systems, and hardware components. The intricate dependencies among these components make version issues hard to detect and resolve and a significant cause of failed builds in DL projects~\cite{han2020empirical}.

{Several techniques~\cite{ye2022knowledge, horton2019dockerizeme, mukherjee2021fixing, wang2021restoring, wang2020watchman} have been proposed to detect dependency issues. However, they primarily focus on detecting dependency issues among Python packages, with limited consideration for issues related to drivers, operating systems, and hardware components~\cite{wang2020watchman}. Moreover, these techniques heavily rely on documented version constraints and dependencies specified in PyPI and official API documentation~\cite{wang2021restoring}, often overlooking undocumented issues that developers encounter in practice. In contrast, Q\&A platforms such as Stack Overflow (SO) offer up-to-date and comprehensive information on dependency issues and their solutions in real-world scenarios. This wealth of version knowledge has not been explored by previous work.}

{In this paper, we present \webtool{}, an interactive web tool that visualizes a knowledge graph containing 2,376 version knowledge extracted from SO discussions. Here, version knowledge refers to (in)compatibility relationships between any two DL stack components (e.g., \texttt{Python 3.7} is compatible with \texttt{TensorFlow 1.5.0}).
These relationships are extracted from 355,000 SO posts that contain version knowledge via UnifiedQA~\cite{khashabi2020unifiedqa}. The extracted knowledge was further consolidated into a knowledge graph among 48 popular DL stack components in \webtool{}. 
Our evaluation of 343 version (in)compatibility relations shows relations in the knowledge graph are highly accurate (83.7\% of the sampled relations are correct).}

{Specifically, \webtool{} provides three major functionalities:}

\begin{enumerate}
    \item \textbf{Visualization}. \webtool{} visualizes a knowledge graph with (in)compatibility relations between DL components.
    \item \textbf{Search Features}. \webtool{} utilizes the function calling ability of GPT-4 to parse natural language search queries of the users, enabling smooth access to any components or relations.
    \item \textbf{Citations}. \webtool{} cites source SO posts, allowing users to validate the extracted knowledge.
\end{enumerate}

 {In our technical paper~\cite{zhao2023knowledge}, we showed this knowledge graph can be used to effectively detect version issues in DL projects. For experiment results, the code of \webtool{}, and more information on our technical paper, please check \url{https://github.com/LexieZhou/Decide}.}



\vspace{2mm}
\begin{figure*}[]
    \centering
    \includegraphics[width=.65\linewidth]{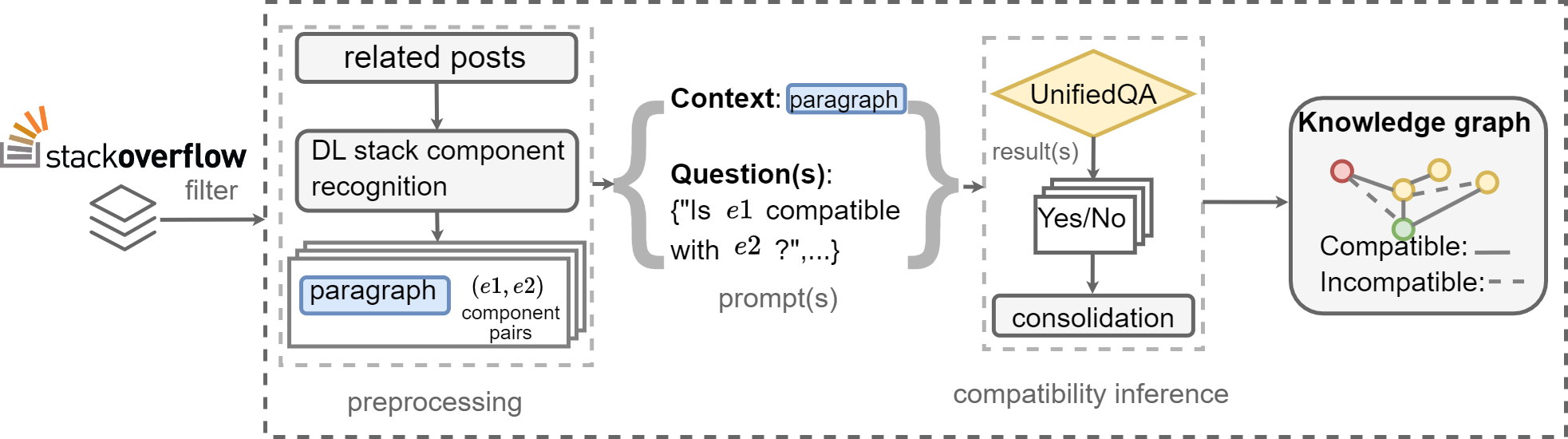}

    \caption{An overview of the knowledge graph construction and incompatibility detection process}
    \label{fig:supervised}
\end{figure*}

\begin{table}[]

\begin{tabularx}{0.45\textwidth}{X}
\hline
\multicolumn{1}{|X|}{\textbf{Context:} \texttt{tensorflow} \texttt{1.13} doesn't work with \texttt{cuda 10.1} because of the following: \textit{``ImportError: libcublas.so.10.0: cannot open shared object file: No such file or directory''}. \texttt{tensorflow} is looking for \texttt{libcublas.so.10.0} whereas \texttt{cuda} provides \texttt{libcublas.so.10.1.0.105}.

\textbf{Question:} 
\hlgreen{Does \texttt{tensorflow} \texttt{1.13} work with \texttt{cuda} \texttt{10.1}?}
}  \\\hline
\end{tabularx}
\vspace{0.2em}
\begin{tabularx}{0.45\textwidth}{X}
\hline
\multicolumn{1}{|X|}{\textbf{Answer from UnifiedQA:} No.}\\ \hline
\end{tabularx}

\captionof{figure}{QA examples for version compatibility inference}
\label{table:qa_example}
\end{table}

\begin{table}[!h]
\caption{Version matching patterns}
\vspace{-6pt}
\label{table:version regex}
\resizebox{\linewidth}{!}{
\SetTblrInner{rowsep=0pt}
\begin{tblr}{
    colspec={Q[l]|l},
}
\hline
\SetCell[c=1]{c}\textsf{Regex Pattern} & \SetCell[c=1]{c}\textsf{Matched Examples} \\
\hline
\SetCell[c=1]{l}\texttt{v\{0,1\}\textbackslash{}d+(\textbackslash.d+)\{1,2\}} & 
\SetCell[c=1]{l}\texttt{3.7, 2.4.3, v2.3, v1.13.5} \\ 
\SetCell[c=1]{l}\texttt{v\{0,1\}\textbackslash d+(\textbackslash.\textbackslash d+)\{0,1\}(\textbackslash.x)\{0,1\}} &
\SetCell[c=1]{l}\texttt{3.x, 1.3.x, v1.x, v2.2.x} \\ 
\SetCell[c=1]{l}\texttt{(COMPONENT)(-| |\_{})}\texttt{v\{0,1\}\textbackslash d+} &
\SetCell[c=1]{l}\texttt{python v3, cuda-8, Windows 64} \\ 
\hline
\end{tblr}
}
\end{table}

{\section{Knowledge Graph Construction}

This section presents how {\tool} extracts version compatibility knowledge from SO posts to build a knowledge graph. Figure~\ref{fig:supervised} provides an overview of the approach. Please refer to our technical paper~\cite{zhao2023knowledge} for more details.

\subsection{Data Collection and Filtering}
We downloaded the Stack Exchange Data Dump~\cite{stackexchange} with 53 million Stack Overflow posts from July 31, 2008 to September 5, 2021. We manually analyzed 798 popular SO tags to identify 46 DL-related tags. \footnote{The complete list of tags can be found at \url{https://github.com/KKZ20/DECIDE/blob/main/DECIDE/docs/SO_tags.json}} We filtered the SO posts to only retain posts tagged with at least one of these tags. 4.9M posts remained after this step. 
Further refinement using 22 linguistic patterns summarized from 150 posts with version incompatibility knowledge\footnote{The complete list of linguistic patterns can be found in the supplementary material.} related to version compatibility issues narrowed the selection to 549K posts. Finally, we removed unaccepted answer posts to ensure the quality of our dataset. After this step, 355K posts remain. A random sample of 384 posts (CI=95\%) showed that 84.9\% of them contain version incompatibility knowledge.

\subsection{DL Stack Component Recognition}
Not all paragraphs in a version-related SO post mentioned version compatibility information. We designed a filtering mechanism to locate paragraphs containing version compatibility information to improve knowledge extraction efficiency.{\tool} only selected paragraphs that mention at least two different versioned components. 
In the current implementation, {\tool} supports the recognition of 48 popular components\footnote{A complete list of 48 DL stack components can be found at  \url{https://github.com/KKZ20/DECIDE/blob/main/DECIDE/docs/DL_Stack_Components.txt}} across five DL stack layers. These components were manually identified from all DL components appearing in the 200 posts with the highest vote score (i.e., upvotes minus downvotes). The authors also added synonyms or aliases for these components to improve recognition accuracy.
 
Furthermore, we designed three regex patterns to identify versions mentioned in a paragraph, as shown in table~\ref{table:version regex}}. Finally, \tool{} matches each component with the closest version in the dependency tree of a sentence using a weighted stable matching algorithm~\cite{wiki:stablemarriage}.

\subsection{Compatibility Inference via a Pre-trained QA Model}
{\tool} infers the compatibility relationship between components based on the paragraph information. We propose a novel approach to reframe the relationship classification task as a Question-Answering (QA) task and use UnifiedQA~\cite{khashabi2020unifiedqa} to infer the compatibility between components. UnifiedQA is a large model with 3 billion parameters pre-trained on eight datasets. It has been demonstrated to understand deep semantics in natural language and achieve state-of-the-art performance in multiple QA benchmarks~\cite{khashabi2020unifiedqa}. UnifiedQA takes two inputs---a {\em question} and a {\em context document} from which the answer is extracted. \tool{} uses the paragraph as a context document and asks UnifiedQA a yes-or-no question to infer the compatibility between the two components. 
Figure~\ref{table:qa_example} illustrates a QA example from the real SO post---\href{https://stackoverflow.com/questions/55028552}{\hlgray{[Post 55028552]}}. 
Considering prompt design has a noticeable impact on model performance~\cite{radford2021learning}, we experimented with eight question templates designed based on the 22 linguistic patterns identified in the post-filtering procedure and found the best prompt\footnote{Both the prompts and our experiment results can be found at  \url{https://github.com/KKZ20/DECIDE/blob/main/DECIDE/docs/DL_Stack_Components.txt}}. If conflicts exist among SO posts, \tool{} chooses the relationship supported by the most posts.

To evaluate the accuracy, we randomly sampled 343 relations from a total of 2,376. For each relation, two authors independently verified whether the relation is true by searching online or performing experiments. Then, they compared their verification results and resolved any disagreement. The Cohen’s Kappa score was 0.89. After independent verification by two authors, 287 relations were confirmed as correct, resulting in an overall accuracy of 83.7\%.



\begin{figure*}
  \centering
  \includegraphics[width=\textwidth]{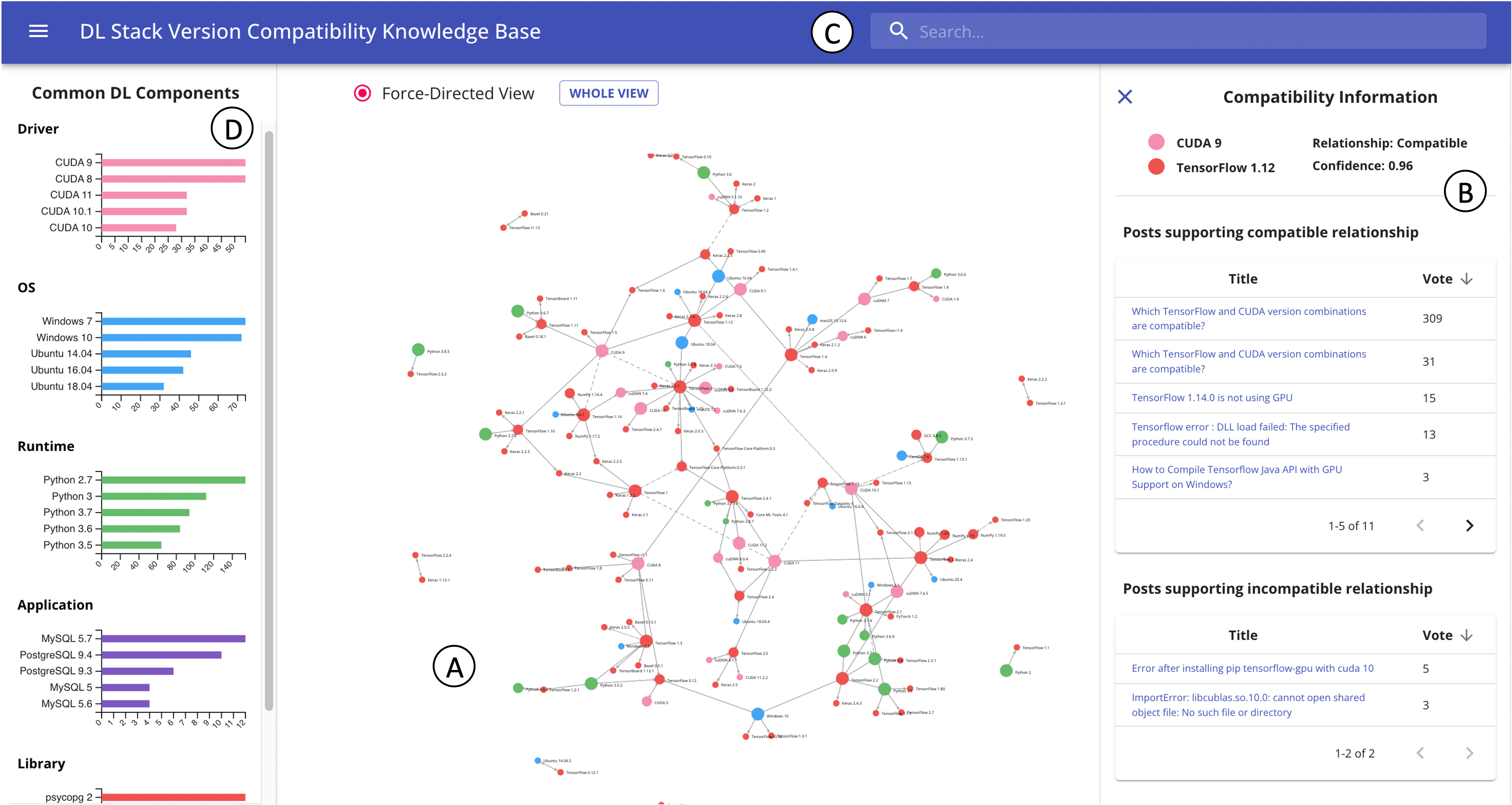}
  \caption{\webtool{}, an interactive knowledge-based tool for detecting and identifying version (in)compatibility in deep learning stacks. (A) The Compatibility Visualizer. (B) The Information Panel. (C) The Search Bar. (D) The Statistical Panel.}
  \label{fig:overview}
\end{figure*}

\section{Knowledge Graph Visualization}

\webtool{} is built with \texttt{React.js} for web application and \texttt{Node.js} for server. \webtool{} consists of four key components (Figure~\ref{fig:overview}): (1) the Compatibility Visualizer (A), (2) the Information Panel (B), (3) the Search Bar (C), and (4) the Statistical Panel (D).

\subsection{The Compatibility Visualizer}
The Compatibility Visualizer (Figure~\ref{fig:overview} A) shows the knowledge graph that visualizes 2,376 compatibility relationships between DL components. 

\textbf{Knowledge Graph} is defined as $G = \langle N, L \rangle$, where:
\begin{align*}
N &= \{n_i \mid i \geq 0, \forall n_i \in \{l, r, d, c, h\}\} \\
L &= \{l_j \mid j \geq 0, \forall l_j \in \{comp, incomp\}\}
\end{align*}

$N$ is the set of nodes denoting DL components with their version numbers. \webtool{} supports the recognition of DL components across five different DL stack layers~\cite{huang2022demystifying}: (1) \textit{library layer} ($l$) that contains the popular frameworks (e.g., PyTorch) and other libraries (e.g., NumPy, SciPy), (2) \textit{runtime layer} ($r$) that contains the execution interpreters or virtual machines of programming languages (e.g., JVM), (3) \textit{driver layer} ($d$) that includes hardware drivers and accelerated SDKs (e.g., CUDA, cuDNN), (4) \textit{OS/container layer}($c$) that includes the operating systems and other containers or virtual environments (e.g., Anaconda, Docker), (5) \textit{hardware layer} ($h$) that includes the hardware and chips (e.g., CPU, GPU).

$L$ is the set of links representing the compatible ($comp$) or incompatible ($incomp$) relationship between two components, denoted by the symbols $\leftrightarrow$ and $\dashleftrightarrow$ respectively. For a pair of versioned components, let $\#\textit{Compatible}$ represent the number of posts that \webtool{} infers a compatible relationship between them, and $\#\textit{Incompatible}$ denote the number of posts that \webtool{} infers an incompatible relationship. We define the confidence score of the relationship between two versioned components as follows: $\textit{confidence score} = \frac{{\#\textit{Compatible} - \#\textit{Incompatible}}}{{\#\textit{Compatible} + \#\textit{Incompatible}}}$. If the \textit{confidence score} is a positive number, it implies a compatible relationship. Otherwise, it implies an incompatible relationship. Relationships with a neutral confidence score ($\textit{confidence score} = 0$) are discarded.

Once we had acquired version knowledge from SO posts, we restructured the version knowledge data into node data ($N$) and link data ($L$) and utilized the JavaScript library \texttt{D3.js}~\footnote{Details about \texttt{D3.js} can be found at \url{https://d3js.org/}}to construct the knowledge graph in the Compatibility Visualizer. 

\subsection{The Information Panel}
The Information Panel (Figure~\ref{fig:overview} B) displays detailed information regarding the DL components themselves as well as the (in)compatibility relationships between them.

\subsubsection{DL Library Statistics}
We extracted all the statistics about DL libraries from \texttt{Libraries.io}, a website that provides comprehensive library information. This retrieved information~\footnote{The retrieved stats data can be found at \url{https://github.com/LexieZhou/Decide/blob/main/stats/stats.json}} includes details such as keywords, licenses, and dependencies. When a node is clicked, \webtool{} retrieves the corresponding component. If the component is a DL library, \webtool{} will display detailed information using the extracted statistics.

\subsubsection{Relationship Details}
When a link is clicked, the information panel will show details of the (in)compatibility relationship between the two DL components. 
Upon clicking, \webtool{} retrieves the link data and displays their predicted (in)compatibility, the confidence score, all the relevant SO posts, and votes of posts. Additionally, we utilize the post IDs stored in the link data to associate the post title and URL further. This allows users to obtain an overview of the posts and access the original SO discussions for further exploration.

\subsection{The Search Bar}

The search bar (Figure~\ref{fig:overview} C) provides users three query options to interact with the knowledge graph. When a query is received, \webtool{} applies filters to the node data and link data, resulting in a smaller dataset that includes relevant nodes and links. The filtered data is then used to generate a focused and concise knowledge graph for exploration.

The first query option allows users to quickly check the compatibility between two different DL components using natural language queries. For example, a user can inquire, \textit{``Does \texttt{Python 3.6.8} work with \texttt{Ubuntu 16.04.6}?''}. To facilitate this process, \webtool{} employs the function calling ability of GPT-4 or regular expression pattern matching (Table~\ref{table:version regex}) to identify the two versioned components. \webtool{} then filters the node data and link data. Subsequently, the knowledge graph shows only the mentioned components and their (in)compatibility relationships. 

Second, users can search for a versioned DL component to see all the associated version knowledge. Using keyword matching, \webtool{} identifies the versioned component in the search query. Subsequently, \webtool{} searches through the available node data to gather all possible versioned nodes matching the search criteria, which are then presented in the search results, allowing users to select. Once users confirm their search selection, \webtool{} further filters the knowledge graph to include the selected component, its connected DL components, and their respective (in)compatible relationships. By doing so, \webtool{} provides users with a comprehensive view of the version knowledge surrounding the chosen component.
For example, searching for \texttt{Python 3.5} will display \texttt{Python 3.5} and its related (in)compatible DL components.

Third, users can search for a DL stack component without a specified version. \webtool{} will show all versions of that component and their version knowledge. Similarly, \webtool{} uses keyword matching to identify the mentioned component and further queries the knowledge graph.

\subsection{The Statistical Panel}
To provide easy access, \webtool{} also includes the Statistical Panel (Figure~\ref{fig:overview} D), which displays the five most popular components with common version issues from each DL stack layer. We calculated the number of version knowledge for each component and identified the top five most discussed components in each DL stack layer. Since version issues about these components are very common, we create shortcuts for these components in the statistical panel.

\section{Evaluation}
\begin{table}
\caption{Accuracy of version incompatibility detection}
\label{tab:evaluation}
\centering
\resizebox{0.3\textwidth}{!}{
\begin{tblr}{width=0.8\textwidth}
\hline
   & \textbf{Precision}&\textbf{Recall}&\textbf{F1 Score}\\
\hline
Watchman~\cite{wang2020watchman} & 16.7\% & 5.9\% & 8.7\%     \\
PyEGo~\cite{ye2022knowledge} & 33.3\% & 29.4\% & 31.2\%      \\
\webtool{} & 91.7\% & 64.7\% & 75.9\% \\
\hline
\end{tblr}
}
\end{table}

Our technical research paper~\cite{zhao2023knowledge} shows \webtool{} can be used to detect version issues in real DL projects by creating a benchmark consisting of 10 popular projects from GitHub.
We first searched for DL projects on GitHub with at least 100 stars and a \textit{requirements.txt} file. We manually reproduced them on our local machine until we found ten projects with at least one version issue on our local machine.~\footnote{The 10 benchmark project statistics can be found at \url{https://github.com/LexieZhou/Decide/blob/main/stats/RQ1/project_statistics.md}} For each experiment, we manually resolved the incompatibility issue. Among the total 17 issues, 4 issues involved components at the same layer, while 13 issues involved components between different DL stack layers.

We compared \webtool{} against two state-of-the-art approaches, PyEGo~\cite{ye2022knowledge} and Watchman~\cite{wang2020watchman}. Table~\ref{tab:evaluation} shows the precison, recall, and F1 score of \webtool{}, Watchman, and PyEGo. Overall, Decide achieves 91.7\% precision and 64.7\% recall, significantly outperforming Watchman and PyEgo. The comparison method detail can be found in our technical paper~\cite{zhao2023knowledge}.

\section{Discussion}
\textbf{Threats to validity.} 
The DL component recognition algorithm employed in our research introduces potential identification errors into our dataset. Additionally, the performance of \webtool{} is dependent on the accuracy and limitations of UnifiedQA. Another potential threat arises from the relatively small benchmark that we used to evaluate \webtool{}. Due to the large volume of SO posts, it was not feasible to manually validate each post for accuracy. Instead, we inspected random samples, which may lead to imprecise estimations.
\\
\textbf{Limitations.} \webtool{} only extracted knowledge from a limited scope from SO, which may limit its effectiveness to only those scenarios and incompatibilities discussed on SO without adapting to new updates. Also, the current version of \webtool{} focuses only on version issues of 48 Python-based DL components. Future research endeavors can expand the knowledge graph by incorporating knowledge from other online documents or apply it to other tasks by adding more diverse SO posts.

\section{Conclusion}
Deep learning has found applications in diverse domains, but version incompatibility issues often lead to build failures in DL projects. In this paper, we introduced \webtool{}, an interactive web-based visualization of a knowledge graph containing 2,376 version knowledge extracted from Stack Overflow discussions. \webtool{} empowers users to confidently reuse or deploy deep learning projects on their local machines, minimizing the likelihood of version-related failures. 


\clearpage
\bibliographystyle{ACM-Reference-Format}
\bibliography{References}


\begin{thebibliography}{15}


\ifx \showCODEN    \undefined \def \showCODEN     #1{\unskip}     \fi
\ifx \showDOI      \undefined \def \showDOI       #1{#1}\fi
\ifx \showISBNx    \undefined \def \showISBNx     #1{\unskip}     \fi
\ifx \showISBNxiii \undefined \def \showISBNxiii  #1{\unskip}     \fi
\ifx \showISSN     \undefined \def \showISSN      #1{\unskip}     \fi
\ifx \showLCCN     \undefined \def \showLCCN      #1{\unskip}     \fi
\ifx \shownote     \undefined \def \shownote      #1{#1}          \fi
\ifx \showarticletitle \undefined \def \showarticletitle #1{#1}   \fi
\ifx \showURL      \undefined \def \showURL       {\relax}        \fi
\providecommand\bibfield[2]{#2}
\providecommand\bibinfo[2]{#2}
\providecommand\natexlab[1]{#1}
\providecommand\showeprint[2][]{arXiv:#2}

\bibitem[sta(2022)]%
        {stackexchange}
 \bibinfo{year}{2022}\natexlab{}.
\newblock \bibinfo{title}{Stack Exchange Data Dump}.
\newblock \bibinfo{howpublished}{Accessed on June 09, 2022}.
\newblock
\urldef\tempurl%
\url{https://archive.org/details/stackexchange}
\showURL{%
\tempurl}


\bibitem[Han et~al\mbox{.}(2020)]%
        {han2020empirical}
\bibfield{author}{\bibinfo{person}{Junxiao Han}, \bibinfo{person}{Shuiguang Deng}, \bibinfo{person}{David Lo}, \bibinfo{person}{Chen Zhi}, \bibinfo{person}{Jianwei Yin}, {and} \bibinfo{person}{Xin Xia}.} \bibinfo{year}{2020}\natexlab{}.
\newblock \showarticletitle{An empirical study of the dependency networks of deep learning libraries}. In \bibinfo{booktitle}{\emph{2020 IEEE International Conference on Software Maintenance and Evolution (ICSME)}}. IEEE, \bibinfo{pages}{868--878}.
\newblock
\urldef\tempurl%
\url{https://doi.org/10.1109/ICSME46990.2020.00116}
\showDOI{\tempurl}


\bibitem[Horton and Parnin(2019)]%
        {horton2019dockerizeme}
\bibfield{author}{\bibinfo{person}{Eric Horton} {and} \bibinfo{person}{Chris Parnin}.} \bibinfo{year}{2019}\natexlab{}.
\newblock \showarticletitle{DockerizeMe: automatic inference of environment dependencies for python code snippets}. In \bibinfo{booktitle}{\emph{Proceedings of the 41st International Conference on Software Engineering}} (Montreal, Quebec, Canada) \emph{(\bibinfo{series}{ICSE '19})}. \bibinfo{publisher}{IEEE Press}, \bibinfo{pages}{328–338}.
\newblock
\urldef\tempurl%
\url{https://doi.org/10.1109/ICSE.2019.00047}
\showDOI{\tempurl}


\bibitem[Huang et~al\mbox{.}(2022)]%
        {huang2022demystifying}
\bibfield{author}{\bibinfo{person}{Kaifeng Huang}, \bibinfo{person}{Bihuan Chen}, \bibinfo{person}{Susheng Wu}, \bibinfo{person}{Junmin Cao}, \bibinfo{person}{Lei Ma}, {and} \bibinfo{person}{Xin Peng}.} \bibinfo{year}{2022}\natexlab{}.
\newblock \showarticletitle{Demystifying dependency bugs in deep learning stack}.
\newblock \bibinfo{journal}{\emph{arXiv preprint arXiv:2207.10347}} (\bibinfo{year}{2022}).
\newblock
\urldef\tempurl%
\url{https://doi.org/10.48550/arXiv.2207.10347}
\showDOI{\tempurl}


\bibitem[Khashabi et~al\mbox{.}(2020)]%
        {khashabi2020unifiedqa}
\bibfield{author}{\bibinfo{person}{Daniel Khashabi}, \bibinfo{person}{Sewon Min}, \bibinfo{person}{Tushar Khot}, \bibinfo{person}{Ashish Sabharwal}, \bibinfo{person}{Oyvind Tafjord}, \bibinfo{person}{Peter Clark}, {and} \bibinfo{person}{Hannaneh Hajishirzi}.} \bibinfo{year}{2020}\natexlab{}.
\newblock \bibinfo{title}{UnifiedQA: Crossing Format Boundaries With a Single QA System}.
\newblock
\newblock
\urldef\tempurl%
\url{https://doi.org/10.48550/arXiv.2005.00700}
\showDOI{\tempurl}
\showeprint[arxiv]{2005.00700}~[cs.CL]


\bibitem[Mukherjee et~al\mbox{.}(2021)]%
        {mukherjee2021fixing}
\bibfield{author}{\bibinfo{person}{Suchita Mukherjee}, \bibinfo{person}{Abigail Almanza}, {and} \bibinfo{person}{Cindy Rubio-Gonz{\'a}lez}.} \bibinfo{year}{2021}\natexlab{}.
\newblock \showarticletitle{Fixing dependency errors for Python build reproducibility}. In \bibinfo{booktitle}{\emph{Proceedings of the 30th ACM SIGSOFT international symposium on software testing and analysis}}. \bibinfo{pages}{439--451}.
\newblock
\urldef\tempurl%
\url{https://doi.org/10.1145/3460319.3464797}
\showDOI{\tempurl}


\bibitem[Radford et~al\mbox{.}(2021)]%
        {radford2021learning}
\bibfield{author}{\bibinfo{person}{Alec Radford}, \bibinfo{person}{Jong~Wook Kim}, \bibinfo{person}{Chris Hallacy}, \bibinfo{person}{Aditya Ramesh}, \bibinfo{person}{Gabriel Goh}, \bibinfo{person}{Sandhini Agarwal}, \bibinfo{person}{Girish Sastry}, \bibinfo{person}{Amanda Askell}, \bibinfo{person}{Pamela Mishkin}, \bibinfo{person}{Jack Clark}, {et~al\mbox{.}}} \bibinfo{year}{2021}\natexlab{}.
\newblock \showarticletitle{Learning transferable visual models from natural language supervision}. In \bibinfo{booktitle}{\emph{International Conference on Machine Learning}}. PMLR, \bibinfo{pages}{8748--8763}.
\newblock
\urldef\tempurl%
\url{https://doi.org/10.48550/arXiv.2103.00020}
\showDOI{\tempurl}


\bibitem[Tian et~al\mbox{.}(2018)]%
        {tian2018deeptest}
\bibfield{author}{\bibinfo{person}{Yuchi Tian}, \bibinfo{person}{Kexin Pei}, \bibinfo{person}{Suman Jana}, {and} \bibinfo{person}{Baishakhi Ray}.} \bibinfo{year}{2018}\natexlab{}.
\newblock \showarticletitle{Deeptest: Automated testing of deep-neural-network-driven autonomous cars}. In \bibinfo{booktitle}{\emph{Proceedings of the 40th international conference on software engineering}}. \bibinfo{pages}{303--314}.
\newblock
\urldef\tempurl%
\url{https://doi.org/10.1145/3180155.3180220}
\showDOI{\tempurl}


\bibitem[Wang et~al\mbox{.}(2021)]%
        {wang2021restoring}
\bibfield{author}{\bibinfo{person}{Jiawei Wang}, \bibinfo{person}{Li Li}, {and} \bibinfo{person}{Andreas Zeller}.} \bibinfo{year}{2021}\natexlab{}.
\newblock \bibinfo{title}{Restoring execution environments of jupyter notebooks. In 2021 IEEE/ACM 43rd International Conference on Software Engineering (ICSE)}.
\newblock
\newblock
\urldef\tempurl%
\url{https://doi.org/10.48550/arXiv.2103.02959}
\showDOI{\tempurl}


\bibitem[Wang et~al\mbox{.}(2020)]%
        {wang2020watchman}
\bibfield{author}{\bibinfo{person}{Ying Wang}, \bibinfo{person}{Ming Wen}, \bibinfo{person}{Yepang Liu}, \bibinfo{person}{Yibo Wang}, \bibinfo{person}{Zhenming Li}, \bibinfo{person}{Chao Wang}, \bibinfo{person}{Hai Yu}, \bibinfo{person}{Shing-Chi Cheung}, \bibinfo{person}{Chang Xu}, {and} \bibinfo{person}{Zhiliang Zhu}.} \bibinfo{year}{2020}\natexlab{}.
\newblock \showarticletitle{Watchman: Monitoring dependency conflicts for python library ecosystem}. In \bibinfo{booktitle}{\emph{Proceedings of the ACM/IEEE 42nd International Conference on Software Engineering}}. \bibinfo{pages}{125--135}.
\newblock
\urldef\tempurl%
\url{https://doi.org/10.1145/3377811.3380426}
\showDOI{\tempurl}


\bibitem[{Wikipedia contributors}(2023)]%
        {wiki:stablemarriage}
\bibfield{author}{\bibinfo{person}{{Wikipedia contributors}}.} \bibinfo{year}{2023}\natexlab{}.
\newblock \bibinfo{title}{Stable marriage problem}.
\newblock \bibinfo{howpublished}{Accessed on January 10, 2024}.
\newblock
\urldef\tempurl%
\url{https://en.wikipedia.org/wiki/Stable_marriage_problem}
\showURL{%
\tempurl}


\bibitem[Wu et~al\mbox{.}(2017)]%
        {wu2017application}
\bibfield{author}{\bibinfo{person}{Qing Wu}, \bibinfo{person}{Yungang Liu}, \bibinfo{person}{Qiang Li}, \bibinfo{person}{Shaoli Jin}, {and} \bibinfo{person}{Fengzhong Li}.} \bibinfo{year}{2017}\natexlab{}.
\newblock \showarticletitle{The application of deep learning in computer vision}. In \bibinfo{booktitle}{\emph{2017 Chinese Automation Congress (CAC)}}. IEEE, \bibinfo{pages}{6522--6527}.
\newblock
\urldef\tempurl%
\url{https://doi.org/10.1109/CAC.2017.8243952}
\showDOI{\tempurl}


\bibitem[Ye et~al\mbox{.}(2022)]%
        {ye2022knowledge}
\bibfield{author}{\bibinfo{person}{Hongjie Ye}, \bibinfo{person}{Wei Chen}, \bibinfo{person}{Wensheng Dou}, \bibinfo{person}{Guoquan Wu}, {and} \bibinfo{person}{Jun Wei}.} \bibinfo{year}{2022}\natexlab{}.
\newblock \showarticletitle{Knowledge-based environment dependency inference for Python programs}. In \bibinfo{booktitle}{\emph{Proceedings of the 44th International Conference on Software Engineering}}. \bibinfo{pages}{1245--1256}.
\newblock
\urldef\tempurl%
\url{https://doi.org/10.1145/3510003.3510127}
\showDOI{\tempurl}


\bibitem[Young et~al\mbox{.}(2018)]%
        {young2018recent}
\bibfield{author}{\bibinfo{person}{Tom Young}, \bibinfo{person}{Devamanyu Hazarika}, \bibinfo{person}{Soujanya Poria}, {and} \bibinfo{person}{Erik Cambria}.} \bibinfo{year}{2018}\natexlab{}.
\newblock \showarticletitle{Recent trends in deep learning based natural language processing}.
\newblock \bibinfo{journal}{\emph{ieee Computational intelligenCe magazine}} \bibinfo{volume}{13}, \bibinfo{number}{3} (\bibinfo{year}{2018}), \bibinfo{pages}{55--75}.
\newblock
\urldef\tempurl%
\url{https://doi.org/10.1109/MCI.2018.2840738}
\showDOI{\tempurl}


\bibitem[Zhao et~al\mbox{.}(2023)]%
        {zhao2023knowledge}
\bibfield{author}{\bibinfo{person}{Zhongkai Zhao}, \bibinfo{person}{Bonan Kou}, \bibinfo{person}{Mohamed~Yilmaz Ibrahim}, \bibinfo{person}{Muhao Chen}, {and} \bibinfo{person}{Tianyi Zhang}.} \bibinfo{year}{2023}\natexlab{}.
\newblock \showarticletitle{Knowledge-Based Version Incompatibility Detection for Deep Learning}.
\newblock \bibinfo{journal}{\emph{arXiv preprint arXiv:2308.13276}} (\bibinfo{year}{2023}).
\newblock
\urldef\tempurl%
\url{https://doi.org/10.1145/3611643.3616364}
\showDOI{\tempurl}


\end{thebibliography}


\end{document}